\documentstyle[12pt,epsf]{article}
\newcommand{\bea}{\begin{eqnarray}}
\newcommand{\eea}{\end{eqnarray}}
\newcommand{\be}{\begin{equation}}
\newcommand{\ee}{\end{equation}}

\newcommand{\nn}{\nonumber}
\newcommand{\rf}[1]{(\ref{#1})}

\begin{document}


\begin{center}
{\bf Model-Independent Confirmation of the $\sigma$-Meson below 1 GeV and
Indication for the $f_0(1500)$ Glueball} \\
\bigskip

Yu.S.~Surovtsev,\\
{\it Bogoliubov Laboratory of Theoretical Physics, Joint Institute for Nuclear
Research, Dubna 141 980, Moscow Region, Russia}\\
D.~Krupa and M.~Nagy\\
{\it Institute of Physics, Slov.Acad.Sci., D\'ubravsk\'a cesta 9,
842 28 Bratislava, Slovakia}
\end{center}
\begin{abstract}
In the model-independent approach consisting in the immediate application to
the analysis of experimental data of such general principles as analyticity
and unitarity, a confirmation of the $\sigma$-meson at $\sim$ 665 MeV and an
indication for the glueball nature of the $f_0(1500)$ state are obtained on
the basis of a simultaneous description of the isoscalar $s$-wave channel of
the $\pi\pi$ scattering (from the threshold up to 1.9 GeV) and of the
$\pi\pi\to K\overline{K}$ process (from the threshold to $\sim$ 1.4 GeV where
the 2-channel unitarity is valid). A parameterless description of the $\pi\pi$
background is first given by allowance for the left-hand branch-point in the
proper uniformizing variable. It is shown that the large $\pi\pi$-background,
usually obtained, combines, in reality, the influence of the left-hand
branch-point and the contribution of a very wide resonance at $\sim$ 665 MeV.
The coupling constants of the observed states with the $\pi\pi$ and
$K\overline{K}$ systems and lengths of the $\pi\pi$ and $K\overline{K}$
scattering are obtained.
\end{abstract}

\section{Introduction}
A problem of scalar mesons is most troublesome and long-lived in the light
meson spectroscopy. A main difficulty in understanding the scalar-isoscalar
sector seems to be related with the possibly-considerable influence of the
vacuum and such effects as the instanton contributions that are difficult to
take into account. But there is another difficulty related to a strong
model-dependence of information on multichannel states obtained in analyses
based on the specific dynamic models or using an insufficiently-flexible
representation of states ({\it e.g.}, the standard Breit -- Wigner form).
Especially, this concerns scalar mesons due to the most weak kinematic
diminution of their widths. It was observed that the scalar mesons are either
very large or, if narrow, lie near the channel thresholds. Earlier, we have
shown \cite{KMS-nc96} that an inadequate description of multichannel states
gives not only their distorted parameters when analyzing data but also can
cause the fictitious states when one neglects important (even
energetic-closed) channels. In this paper we are going, conversely, to
demostrate that the large background ({\it e.g.}, that happens in analyzing
$\pi\pi$ scattering), can hide low-lying states (even such important for
theory as a $\sigma$-meson \cite{PDG-98}). With this object, a very
interesting and instructive history is related. A majority of analyses
rejected this meson by resolving the known "up-down ambiguity" in the
700-900-MeV region in the solutions of the $\pi\pi$ phase-shift analyses for
$\delta^0_0$ in favour of the "down" one, because to the "up" solution, one
related the $\epsilon(800)$ resonance of width $\sim$ 150-200 MeV. However,
some of theorists continued to insist on the existence of that state because
it is required by most of the models (like the linear $\sigma$-models or the
Nambu -- Jona-Lasinio models \cite{NJL}) for spontaneous breaking of chiral
symmetry. Since all the analyses of the $s$-wave $\pi\pi$ scattering gave a
large $\pi\pi$-background, it was said that this state (if exists) is
"unobservably"-wide. Recently, new analyses of the old and new experimental
data have been performed which give a very wide scalar-isoscalar state in the
energy region 500-850 MeV \cite{Zou}-\cite{Kaminski}. However, these analyses
use either the Breit -- Wigner form (even if modified) or specific forms of
interactions in a quark model, unitarized by taking the relevant process
thresholds into account, or in a multichannel approach to the considered
processes; therefore, there one cannot talk about a model independence of
results. Besides, in these analyses, a large $\pi\pi$-background is obtained.
We are going to show that a proper detailing of the background (as allowance
for the left-hand branch-point) permits us to extract from the latter a very
wide (but observable) state below 1 GeV even in the "down" solution for the
$\pi\pi$ phase-shift, and, therefore, it is highly important in studying
lightest states.

An adequate consideration of multichannel states and a model-independent
information on them can be obtained on the basis of the first principles
(analyticity, unitarity and Lorentz invariance) immediately applied to
analyzing experimental data. The way of realization is a consistent allowance
for the nearest singularities on all sheets of the Riemann surface of the
$S$-matrix. The Riemann-surface structure is taken into account by a proper
choice of the uniformizing variable. Earlier, we have proposed this method for
2- and 3-channel resonances and developed the concept of standard clusters
(poles on the Riemann surface) as a qualitative characteristic of a state
and a sufficient condition of its existence as well as a criterion of a
quantitative description of the coupled-process amplitudes when all the
complifications of the analytic structure due to a finite width of resonances
and crossing channels and high-energy ``tails'' are accumulated in quite a
smooth background \cite{KMS-nc96,KMS-cz88,KMS-L86}. Let us stress that for a
wide state, the pole position (the pole cluster one for multichannel states)
is a more stable characteristic than the mass and width which are strongly
dependent on a model. The cluster kind is determined from the analysis of
experimental data and is related to the state nature. At all events, we can,
in a model-independent manner, discriminate between bound states of particles
and the ones of quarks and gluons, qualitatively predetermine the relative
strength of coupling of a state with the considered channels, and obtain an
indication on its gluonium nature.

Since, in this work, a main stress is laid on studying lowest states, it is
sufficient to restrict itself to a two-channel approach when considering
simultaneously the coupled processes $\pi\pi\to \pi\pi,K\overline{K}$.
Furthermore, in the uniformizing variable, one must take into account, besides
the branch-points corresponding to the thresholds of the processes $\pi\pi\to
\pi\pi,K\overline{K}$, also the left-hand branch-point at $s=0$, related to
the background in which the crossing-channel contributions are contained.

The layout of the paper is as follows. In Section 2, we outline the
two-coupled-channel formalism, determine the pole clusters on the Riemann surface
as characteristics of multichannel states, and introduce a new uniformizing
variable, allowing for the branch-points of the right-hand (unitary) and
left-hand cuts of the $\pi\pi$-scattering amplitude. In Section 2, we analyze
simultaneously experimental data on the processes $\pi\pi\to \pi\pi,
K\overline{K}$ in the isoscalar $s$-wave on the basis of the presented
approach. In the Conclusion, the obtained results are discussed.

\section{Two-Coupled-Channel Formalism}

Considering the multichannel problem (here the 2-channel one), we pursue
two aims: to obtain a model-independent information about the multichannel
resonances and an indication about their QCD nature, and to describe the
experimental data on the coupled processes. The first purpose is achieved
through the account of the nearest (to the physical region of interest)
singularities of the $S$-matrix. Herewith it is important to analyze
simultaneously experimental data on the coupled processes.

Here we consider the coupled processes of $\pi\pi$ and $K\overline{K}$
scattering and $\pi\pi\to K\overline{K}$. Therefore, we have the two-channel
$S$-matrix determined on the 4-sheeted Riemann surface. The $S$-matrix
elements $S_{\alpha\beta}$, where $\alpha,\beta=1(\pi\pi), 2(K\overline{K})$,
have the right-hand (unitary) cuts along the real axis of the $s$-variable
complex plane, starting at the points $4m_\pi^2$ and $4m_K^2$ and extending
to $\infty$, and the left-hand cuts, which are related to the crossing-channel
contributions and extend along the real axis towards $-\infty$ and begin at
$s=0$ for $S_{11}$ and at $4(m_K^2-m_\pi^2)$ for $S_{22}$ and $S_{12}$. We
number the Riemann-surface sheets according to the signs of analytic
continuations of the channel momenta
\be \label{k_i}
k_1=(s/4-m_\pi^2)^{1/2}, \qquad k_2=(s/4-m_K^2)^{1/2}
\ee
as follows:  signs $({\mbox{Im}}k_1,{\mbox{Im}}k_2)=++,-+,--,+-$ correspond to
the sheets I,II,III,IV. Then, for instance, from the physical region on sheet
I we pass across the cut below the $K\overline{K}$ threshold to sheet II;
above the $K\overline{K}$ threshold, to sheet III.

To elucidate the resonance representation on the Riemann surface, we express
analytic continuations of the matrix elements to the unphysical sheets
$S_{\alpha\beta}^L$ ($L=II,III,IV$) in terms of them on the physical sheet
$S_{\alpha\beta}^I$. Those expressions are convenient for our purpose because,
on sheet I (the physical sheet), the matrix elements $S_{\alpha\beta}^I$ can
have only zeros beyond the real axis. Using the reality property of the
analytic functions and the 2-channel unitarity, one can obtain
\bea \label{S_L}
&&S_{11}^{II}=\frac{1}{S_{11}^I},\qquad ~~~~S_{11}^{III}=\frac{S_{22}^I}{\det S^I},
\qquad S_{11}^{IV}=\frac{\det S^I}{S_{22}^I},\nn\\
&&S_{22}^{II}=\frac{\det S^I}{S_{11}^I},\qquad S_{22}^{III}=\frac{S_{11}^I}
{\det S^I},\qquad S_{22}^{IV}=\frac{1}{S_{22}^I},\\
&&S_{12}^{II}=\frac{iS_{12}^I}{S_{11}^I},\qquad ~~~S_{12}^{III}=\frac{-S_{12}^I}
{\det S^I},\qquad S_{12}^{IV}=\frac{iS_{12}^I}{S_{22}^I},\nn
\eea
Here $\det S^I=S_{11}^I S_{22}^I-(S_{12}^I)^2$. Provided a resonance has the
only decay mode (1-channel case), in the matrix element, the resonance (in the
limit of its narrow width) is represented by a pair of complex conjugate poles
on the IInd sheet and by a pair of conjugate zeros on the physical sheet at
the same points of complex energy. This model-independent statement about the
poles as the nearest singularities holds also when taking account of the
finite width of a resonance. In the case of two coupled channels, formulae
\rf{S_L} immediately give the resonance representation by poles and zeros on
the 4-sheeted Riemann surface. Here one must discriminate between three types
of resonances -- which are described ({\bf a}) by a pair of complex conjugate
poles on sheet II and, therefore, by a pair of complex conjugate zeros on the
Ist sheet in $S_{11}$; ({\bf b}) by a pair of conjugate poles on sheet IV and,
therefore, by a pair of complex conjugate zeros on sheet I in $S_{22}$;
({\bf c}) by one pair of conjugate poles on each of sheets II and IV, that is
by one pair of conjugate zeros on the physical sheet in each of matrix
elements $S_{11}$ and $S_{22}$.

As is seen from \rf{S_L}, to the resonances of types ({\bf a}) and ({\bf b})
one has to make correspond a pair of complex conjugate poles on sheet III
which are shifted relative to a pair of poles on sheet II and IV ,
respectively (if the coupling among channels were absent, i.e. $S_{12}=0$, the
poles on sheet III would lay exactly ({\bf a}) under the poles on the IInd
sheet, ({\bf b}) above the poles on the IVth sheet). To the states of type
({\bf c}) one must make correspond two pairs of conjugate poles on sheet III
which are reasonably expected to be a pair of the complex conjugate compact
formations of poles. Thus, we arrive at the notion of three standard
pole-clusters which represent two-channel bound states of quarks and gluons.
It is convenient to discriminate between those clusters according to the
presence of zeros, corresponding to the state, on the physical sheet in matrix
element $S_{11}$ ({\bf a}), $S_{22}$ ({\bf b}) or in both ({\bf c}).

Note that this resonance division into types is not formal. In paricular,
the resonance, coupled strongly with the first ($\pi\pi$) channel, is
described by the pole cluster of type ({\bf a}); if the resonance is coupled
strongly with the $K\overline{K}$ and weakly with $\pi\pi$ channel (say, if it
has a dominant $s\overline{s}$ component), then it is represented by the
cluster of type ({\bf b}); finally, since a most noticeable property of a
glueball is the flavour-singlet structure of its wave function and, therefore,
(except the factor $\sqrt{2}$ for a channel with neutral particles)
practically equal coupling with all the members of the nonet, then a glueball
must be represented by the pole cluster of type ({\bf c}) as a necessary
condition.

Just as in the 1-channel case, the existence of a particle bound-state means
the presence of a pole on the real axis under the threshold on the physical
sheet, so in the 2-channel case, the existence of a bound state in channel 2
($K\overline{K}$ molecule), which, however, can decay into channel 1
($\pi\pi$ decay), would imply the presence of a pair of complex conjugate
poles on sheet II under the threshold of the second channel without an
accompaniment of the corresponding shifted pair of poles on sheet III. Namely,
according to this test, earlier an interpretation of the $f_0 (980)$ state as
$K\overline{K}$ molecule has been rejected \cite{KMS-nc96,KMS-cz88,MP-92}.

Generally, formulae \rf{S_L} are a solution of the 2-channel problem in the
sense of giving a chance to predict (on the basis of the data on one process)
the coupled-process amplitudes under a certain conjecture about the
background. We made this earlier in the 2-channel approach \cite{KMS-cz88}.
It was a success to describe ($\chi^2/ndf\approx1.06$) the experimental
isoscalar $s$-wave of $\pi\pi$ scattering from the threshold to 1.9 GeV, to
predict satisfactorily (on the basis of data on $\pi\pi$ scattering) the
behaviour of the $s$-wave of $\pi\pi\to K\overline{K}$ process approximately
up to 1.25 GeV. To take account of the proper right-hand branch-points, the
corresponding uniformizing variable has been used. However, for the
simultaneous analysis of experimental data on the coupled processes it is more
convenient to use the Le Couteur-Newton relations \cite{LN} representing
compactly all features given by formulae \rf{S_L} and expressing the
$S$-matrix elements of all coupled processes in terms of the Jost matrix
determinant $d(k_1,k_2)\equiv d(s)$, the real analytic function with the only
square-root branch-points at the process thresholds $k_i=0$ \cite{Kato}.
Earlier, this was done by us in the 2-channel consideration
\cite{KMS-cz88} with the uniformizing variable
\be \label{z}
z=\frac{k_1+k_2}{\sqrt{m_K^2-m_\pi^2}},
\ee
which was proposed in Ref. \cite{Kato} and maps the 4-sheeted Riemann surface
with two unitary cuts, starting at the points $4m_\pi^2$ and $4m_K^2$, onto
the plane. (Note that other authors have been also applied the
parametrizations with using the Jost functions at analyzing the $s$-wave
$\pi\pi$ scattering in the one-channel approach \cite{Bohacik} and in the
two-channel one \cite{MP-92}. In latter work, the uniformizing variable
$k_2$ has been used, therefore, their approach cannot be emploied near by
the $\pi\pi$ threshold.)

When analyzing the processes $\pi\pi\to \pi\pi,K\overline{K}$ by the above
methods in the 2-channel approach, two resonances ($f_0 (975)$ and
$f_0 (1500)$) are found to be sufficient for a satisfactory description
($\chi^2/\mbox{ndf}\approx1.00$). However, in this case, the large
$\pi\pi$-background has been obtained. A character of the representation of
the background (the pole of second order on the imaginary axis on sheet II and
the corresponding zero on sheet I) suggests that a wide light state is
possible to be hidden in the background. To check this, one must work out the
background in some detail.

Now we will take, in the uniformizing variable, into account also the
left-hand branch-point at $s=0$. We use the uniformizing variable
\be \label{v}
v=\frac{m_K\sqrt{s-4m_\pi^2}+m_\pi\sqrt{s-4m_K^2}}{\sqrt{s(m_K^2-m_\pi^2)}},
\ee
which maps the 4-sheeted Riemann surface, having (in addition to two
above-indicated unitary cuts) also the left-hand cut starting at the point
$s=0$, onto the $v$-plane. It is convenient to write also the function $s(v)$
\be \label{s(v)}
s=-\frac{16m_K^2m_\pi^2v^2}{(m_K^2-m_\pi^2)(v-b)(v+b)(v-b^{-1})(v+b^{-1})},
\ee
where $b=\sqrt{(m_K+m_\pi)/(m_K-m_\pi)}$ is the point into which $s=\infty$
is mapped on the $v$-plane, and the symmetry properties of this function
\be \label{s-v}
s(v)=s(-v)=s(v^{-1})=s(-v^{-1})=s^*(v^*)
\ee
demostrate which points on the $v$-plane correspond to the same point on the
$s$-plane. In Fig.1, the plane of the uniformizing variable $v$ for the
$\pi\pi$-scattering amplitude is depicted. The Roman numerals (I,\ldots,IV)
denote the images of the corresponding sheets of the Riemann surface; the
thick line represents the physical region; the points i, 1 and $b$ correspond
to the $\pi\pi, K\overline{K}$ thresholds and $s=\infty$, respectively; the
shaded intervals $(-\infty,-b],~[-b^{-1},b^{-1}],~[b,\infty)$ are the images
of the corresponding edges of the left-hand cut.
The depicted positions of poles ($*$) and of zeros ($\circ$) give the
representation of the type ({\bf a}) resonance in $S_{11}$. In Fig.1, a very
symmetric picture is shown which ensures the known fact that the $\pi\pi$
interaction is practically elastic up to the $K\overline{K}$ threshold (the
contribution of the multiparticle states ($4\pi,6\pi$) is negligible within
the up-to-date experiment accuracy). This property of the $\pi\pi$ interaction
is satisfied since the poles and zeros are symmetric to each other with
respect to the unit circle. If the $\pi\pi$ scattering were elastic also above
the $K\overline{K}$ threshold, there would be the symmetry of the poles and
zeros with respect to the real axis. The symmetry of the whole picture
relative to the imaginary axis ensures the property of the real analyticity.

On $v$-plane the Le Couteur-Newton relations are \cite{KMS-cz88,Kato}
\be \label{v:C-Newton}
S_{11}=\frac{d(-v^{-1})}{d(v)},\quad S_{22}=\frac{d(v^{-1})}{d(v)},
\quad S_{11}S_{22}-S_{12}^2=\frac{d(-v)}{d(v)}.
\ee
Then, the condition of the real analyticity implies
\be \label{z:real-anal}
d(-v^*)=d^* (v)
\ee
for all $v$, and the unitarity needs the following relations to hold true for
the physical $v$-values:
\be \label{v:unit.requir.}
|d(-v^{-1})|\leq |d(v)|,\quad |d(v^{-1})|\leq |d(v)|,\quad |d(-v)|=|d(v)|.
\ee
The $d$-function that on the $v$-plane already does not possess branch-points is taken as
\be \label{d}
d=d_B d_{res},
\ee
where ~$d_B=B_{\pi}B_K$; $B_{\pi}$ contains the possible remaining
$\pi\pi$-background contribution, related to exchanges in crossing channels;
$B_K$ is that part of the $K\overline{K}$ background which does not contribute
to the $\pi\pi$-scattering amplitude. The most considerable part of the
background of the considered coupled processes related to the influence of the
left-hand branch-point at $s=0$ is taken already in the uniformizing variable
$v$ \rf{v} into account. The function $d_{res}(v)$ represents the contribution
of resonances, described by one of three types of the pole-zero clusters,
{\it i.e.}, except for the point $v=0$, it consists of zeros of clusters:
\be \label{d_res}
d_{res} = v^{-M}\prod_{n=1}^{M} (1-v_n^* v)(1+v_n v),
\ee
where $n$ runs over the independent zeros; therefore, for resonances of the
types ({\bf a}) and ({\bf b}), $n$ has two values, for the type ({\bf c}),
four values; $M$ is the number of pairs of the conjugate zeros.

\section{Analysis of experimental data.}

Using the described 2-channel approach, we analyze simultaneously the
available experimental data on the $\pi\pi$-scattering \cite{Hyams} and
the process $\pi\pi\to K\overline{K}$ \cite{Wickl} in the channel with
$I^GJ^{PC}=0^+0^{++}$. As data, we use the results of phase analyses which
are given for phase shifts of the amplitudes ($\delta_1$ and $\delta_{12}$)
and for moduli of the $S$-matrix elements $\eta_1$ (the elasticity parameter)
and $\xi$:
\be \label{del.eta}
S_a =\eta_a e^{2i\delta_a}~~~~(a=1,2),\qquad S_{12} =i\xi e^{i\delta_{12}}.
\ee
(Remember that "1" denotes here the $\pi\pi$ channel, "2" -- $K\overline{K}$).
The 2-channel unitarity condition gives
\be \label{2-chan.unit.cond.}
\eta_1=\eta_2=\eta,\qquad \xi=(1-\eta^2)^{1/2},\qquad \delta_{12} =
\delta_1+\delta_2.
\ee
We have taken the data on the $\pi\pi$ scattering from the threshold up to
1.89 GeV. Then, comparing experimental data for $\xi$ with values of $\xi$,
calculated by eq.\rf{2-chan.unit.cond.} with using the experimental points
for the elasticity parameter $\eta$, one can see that the 2-channel unitarity
takes place approximately to 1.4 GeV.

To obtain the satisfactory description of the $s$-wave $\pi\pi$ scattering
from the threshold to 1.89 GeV (Fig.2 and Fig.3), we have taken $B_\pi=1$ in
eq.\rf{d}, and three multichannel resonances turned out to be sufficient: the
two ones of the type ({\bf a}) ($f_0 (665)$ and $f_0 (980)$) and $f_0 (1500)$
of the type ({\bf c}). Therefore, in eq.\rf{d_res} $M=8$ and the following
zero positions on the $v$-plane, corresponding to these resonances, have been
established in this situation with the parameterless description of the
background:
\bea
{\rm for} ~~f_0 (665):
~~&&v_1=1.36964+0.208632i,\qquad v_2 =0.921962-0.25348i,\nn\\
{\rm for} ~~f_0 (980):~
&&v_3=1.04834+0.0478652i,\qquad ~v_4 =0.858452-0.0925771i,\nn\\
{\rm for} ~~f_0 (1500):
&&v_5=1.2587+0.0398893i,\qquad ~~~v_6 =1.2323-0.0323298i,\nn\\
&&v_7=0.809818-0.019354i,\qquad v_8 =0.793914-0.0266319i.\nn
\eea
Here for the phase shift $\delta_1$ and the elasticity parameter $\eta$, 113
and 50 experimental points \cite{Hyams}, respectively, are used; when
rejecting the points at energies 0.61, 0.65, and 0.73 GeV for $\delta_1$
and at 0.99, 1.65, and 1.85 GeV for $\eta$, which give an anomalously
large contribution to $\chi^2$, we obtain for $\chi^2/\mbox{ndf}$ the values
2.7 and 0.72, respectively; the total $\chi^2/\mbox{ndf}$ in the case of the
$\pi\pi$ scattering is 1.96.

With the presented picture, the satisfactory description for the modulus
($\xi$) of the $\pi\pi\to K\overline{K}$ matrix element is given from the
threshold to $\sim$ 1.4 GeV (Fig.4). Here 35 experimental points \cite{Wickl}
are used; $\chi^2/\mbox{ndf}\approx 1.11$ when eliminating the points at
energies 1.002, 1.265, and 1.287 GeV (with especially large contribution to
$\chi^2$). However, for the phase shift $\delta_{12}(s)$, slightly excessive
curve is obtained. Therefore, keeping the {\it parameterless} description of
the $\pi\pi$ background, one must take into account the part of the
$K\overline{K}$ background that does not contribute to the
$\pi\pi$-scattering amplitude. Furthermore, this contribution is to be
elastic. Note that the variable $v$ is uniformizing for the
$\pi\pi$-scattering amplitude, {\it i.e.}, on the $v$-plane, $S_{11}$ has no
cuts, however, the amplitudes of the $K\overline{K}$ scattering and
$\pi\pi\to K\overline{K}$ process do have the cuts on the $v$-plane, which
arise from the left-hand cut on the $s$-plane, starting at the point
$s=4(m_K^2-m_\pi^2)$. Under the $s\to v$ conformal mapping \rf{v}, this
left-hand cut is mapped into cuts which begin at the points
$$v=\frac{m_K\sqrt{m_K^2-2m_\pi^2}\pm im_\pi}{m_K^2-m_\pi^2}$$
on the unit circle on the $v$-plane, go along it up to the imaginary axis, and
occupy the latter. This left-hand cut will be neglected in the Riemann-surface
structure, and the contribution on the cut will be taken into account in the
$K\overline{K}$ background as a pole on the real $s$-axis on the physical
sheet in the sub-$K\overline{K}$-threshold region; on the $v$-plane, this pole
gives two poles on the unit circle in the upper half-plane, symmetric to each
other with respect to the imaginary axis, and two zeros, symmetric to the
poles with respect to the real axis, {\it i.e.} at describing the process $\pi\pi\to
K\overline{K}$, one additional parameter is introduced, say, a position $p$ of
the zero on the unit circle. Therefore, for $B_K$ in eq. \rf{d} we take the
form
\be \label{B_K}
B_K=v^{-4}(1-pv)^4(1+p^*v)^4.
\ee
Fourth power in \rf{B_K} is stipulated by the following. First, a pole on the
real $s$-axis on the physical sheet in $S_{22}$ is accompanied by a pole in
sheet II at at the same $s$-value (as it is seen from eqs. \rf{S_L}); on the
$v$-plane, this implies the pole of second order (and also zero of the same
order, symmetric to the pole with respect to the real axis). Second, for the
$s$-channel process $\pi\pi\to K\overline{K}$, the crossimg $u$- and
$t$-channels are the $\pi-K$ and $\overline{\pi}-K$ scattering (exchanges in
these channels give contributions on the left-hand cut); this results in the
additional doubling of the multiplicity of the indicated pole on
the $v$-plane. Zeros of the fourth order in $B_K$ (and, correspondingly, poles
of the fourth order in the $K\overline{K}$-amplitude) provide the better
description of the $K\overline{K}$ background than the ones of the first order
in our recent work \cite{skn-99}.
One can verify that the expression \rf{B_K} does not contribute to $S_{11}$,
{\it i.e.} the parameterless description of the $\pi\pi$ background is kept.
A satisfactory description of the phase shift $\delta_{12}(\sqrt{s})$ (Fig.5)
is obtained approximately to 1.52 GeV with the value of the parameter
$p=0.948201+0.31767i$ (this corresponds to the position of the pole on the
$s$-plane at $s=0.434 {\rm GeV}^2$). Here 59 experimental points \cite{Wickl}
are considered; $\chi^2/\mbox{ndf}\approx 3.05$ when eliminating the points at
energies 1.117, 1.247, and 1.27 GeV (with especially large contribution to
$\chi^2$). The total $\chi^2/\mbox{ndf}$ for four analyzed quantities to
describe the coupled processes $\pi\pi\to\pi\pi,K\overline{K}$ is 2.12; the
number of adjusted parameters is 17, where they all (except a single relating
to the $K\overline{K}$ background) are positions of poles describing
resonances.

In Table 1, the obtained parameter values of poles on the corresponding sheets
of the Riemann surface are cited on the complex energy plane ($\sqrt{s_r}=
{\rm E}_r-i\Gamma_r$). We stress that these are not masses and widths of
resonances. Since, for wide resonances, values of masses and widths are very
model-dependent, it is reasonable to report characteristics of pole clusters
which must be rather stable for various models.
\begin{table}[htb]
\begin{center}
Table 1.
\end{center}
\begin{center}
\begin{tabular}{|c|rl|rl|rl|rl|rl|rl|}
\hline
{} & \multicolumn{2}{c|}{$f_0 (665)$} & \multicolumn{2}{c|}{$f_0(980)$}
& \multicolumn{2}{c|}{$f_0(1500)$} \\
\cline{2-7}
Sheet & \multicolumn{1}{c}{E, MeV} & \multicolumn{1}{c|}{$\Gamma$, MeV}
& \multicolumn{1}{c}{E, MeV} & \multicolumn{1}{c|}{$\Gamma$, MeV}
& \multicolumn{1}{c}{E, MeV} & \multicolumn{1}{c|}{$\Gamma$, MeV}\\
\hline
II & 610$\pm$14 & 620$\pm$26 & 988$\pm$5 & 27$\pm$8 & 1530$\pm$25
& 390$\pm$30 \\
\hline
III & 720$\pm$15 & 55$\pm$9 & 984$\pm$16 & 210$\pm$22 & 1430$\pm$35
& 200$\pm$30 \\
{} & {} & {} & {} & {} & 1510$\pm$22~ & 400$\pm$34 \\
\hline
IV & {} & {} & {} & {} & 1410$\pm$24 & 210$\pm$38  \\
\hline \end{tabular}
\end{center}
\end{table}

Now we can calculate the constants of the obtained-state couplings with the
$\pi\pi-"1"$ and $K\overline{K}-"2"$ systems through the residues of amplitudes at the
pole on sheet II. Expressing the $T$-matrix via the $S$-matrix as
\be \label{T-S}
S_{ii}=1+2i\rho_i T_{ii},\qquad S_{12}=2i\sqrt{\rho_1\rho_2} T_{12},
\ee
where $\rho_i=\sqrt{(s-4m_i^2)/s}$, and taking the resonance part of the
amplitude in the form
\be \label{T_res}
T_{ij}^{res}=\sum_r g_{ir}g_{rj}D_r^{-1}(s),
\ee
where $D_r(s)$ is an inverse propagator ($D_r(s)\propto s-s_r$), we define
the coupling constants as
\be \label{pi-pi-f_0}
g_i g_j=\frac{16m_K^2m_\pi^2}{3(m_K^2-m_\pi^2)}
\left|\frac{{v_r^*}^{2}-{v_r^*}^{-2}}{({v_r^*}^2-b^2)({v_r^*}^2-b^{-2})
({v_r^*}^{-2}-b^2)({v_r^*}^{-2}-b^{-2})}\lim_{v\to {v_r^*}^{-1}}
(1-v_r^*v)\frac{S_{ij}(v)}{\sqrt{\rho_i\rho_j}}\right| .
\ee
Here we denote the coupling constants with the $\pi\pi$ and $K\overline{K}$
systems through $g_1$ and $g_2$, respectively. The obtained values of the
coupling constants of the observed states are given in Table 2.
\begin{table}[htb]
\begin{center} Table 2. \end{center}
\begin{center}
\begin{tabular}{|l|l|l|l|} \hline
{}  & $f_0(665)$ & $f_0(980)$ & $f_0(1500)$\\ \hline
$g_{1}$, GeV & ~$0.7477\pm 0.095$~ & ~$0.1615 \pm 0.03$~ & ~$0.899 \pm 0.093$~\\
\hline
$g_2$, GeV & ~$0.834\pm 0.1$~ & ~$0.438 \pm 0.028$~ & {}\\
\hline
\end{tabular} \end{center}
\end{table}

In this 2-channel approach, there is no point in calculating the coupling
constant of the $f_0(1500)$ state  with the $K\overline{K}$ system, because
the 2-channel unitarity is valid only to 1.4 GeV, and, above this energy,
there is a considerable disagreement between the calculation of the amplitude
modulus $S_{12}$ and the experimental data.

Let us indicate also scattering lengths calculated in our approach. For the
$K\overline{K}$ scattering, we obtain
$$a_0^0(K\overline{K})=-1.188\pm 0.13+(0.648\pm 0.09)i,~m_{\pi^+}^{-1}.$$
A presence of the imaginary part in $a_0^0(K\overline{K})$ reflects the fact,
that already at the threshold of the $K\overline{K}$ scattering, other
channels ($2\pi,4\pi$ etc.) are opened. In Table 3, we have presented our
result for the $\pi\pi$ scattering length $a_0^0$ and its comparison with
results of some other works both theoretical and experimental.
\begin{table}[htb]
\begin{center} Table 3. \end{center}
\begin{center}
\begin{tabular}{|c|l|l|} \hline
$a_0^0, ~m_{\pi^+}^{-1}$ & ~~~~~~~References & ~~~~~~~~~~~~~~~~~Remarks \\
\hline
$0.27\pm 0.06$ & our paper & model-independent approach \\
\hline
$0.26\pm 0.05$ & L. Rosselet et al.\cite{Hyams} & analysis of the decay
$K\to\pi\pi e\nu$ \\
{} & {} & with using Roy's model\\
\hline
$0.24\pm 0.09$ & A.A. Bel'kov et al.\cite{Hyams} & analysis of the process
$\pi^-p\to\pi^+\pi^-n$ \\
{} & {} & with using the effective range formula\\
\hline
$0.23$ & S. Ishida et al.\cite{Ishida} & modified approach to analysis of
$\pi\pi$ scattering \\
{} & {} & with using Breit-Wigner forms \\
\hline
$0.16$ & S. Weinberg \cite{Weinberg} & current algebra (non-linear
$\sigma$-model) \\
\hline
$0.20$ & J. Gasser, H. Leutwyler \cite{Gasser} & chiral theory with
one-loop corrections,\\
{} & {} & non-linear realization of chiral symmetry \\
\hline
$0.217$ & J. Bijnens at al.\cite{Bijnens} & chiral theory with two-loop
corrections,\\
{} & {} & non-linear realization of chiral symmetry  \\
\hline
$0.26$ & M.K. Volkov \cite{Volkov} & chiral theory, \\
{} & {} & linear realization of chiral symmetry \\
\hline
\end{tabular} \end{center}
\end{table}

We have here presented model-independent results: the pole positions,
coupling constants and scattering lengths. The formers can be used further
for calculating masses and widths of these states in various models.

If we suppose, that the obtained state $f_0(665)$ is the $\sigma$-meson, then
from the known relation of the $\sigma$-model between the coupling constant
of the $\sigma$ with the $\pi\pi$-system and masses
$$g_{\sigma\pi\pi}=\frac{m_\sigma^2-m_\pi^2}{\sqrt{2}f_{\pi^0}}$$
(here $f_{\pi^0}$ is the constant of the weak decay of the $\pi^0$:
$f_{\pi^0}=93.1$ MeV), we obtain ~$m_\sigma\approx 342$ MeV. That small value
of the $\sigma$-mass can be a result of the mixing with the $f_0(980)$ state
\cite{Volk-Yud}.

\section{Conclusions}

In the present work, in the model-independent approach consisting in the
immediate application to the analysis of experimental data of first principles
(analyticity-causality and unitarity), a satisfactory simultaneous description
of the isoscalar $s$-wave channel of the processes $\pi\pi\to \pi\pi,
K\overline{K}$ from the thresholds to the energy values, where the 2-channel
unitarity is valid, is obtained. A parameterless description of the $\pi\pi$
background is first given by allowance for the left-hand branch-point in the
proper uniformizing variable. It is shown that the large $\pi\pi$-background,
usually obtained, combines in reality the influence of the left-hand
branch-point and the contribution of a very wide resonance at $\sim$ 665 MeV.
Thus, a model-independent confirmation of the state, already discovered in
other works \cite{Svec}-\cite{Kaminski} (or pretending to this discovery) and
denoted in the PDG issues by $f_0(400-1200)$ \cite{PDG-98}, is obtained.
This is the $\sigma$-meson required by majority of models for spontaneous
breaking of chiral symmetry. Three states ($f_0(665) - \sigma$-meson,
$f_0 (980)$ and $f_0(1500)$) are sufficient to describe the analyzed data.

The discovery of the $f_0(665)$ state solves one important mystery of the
scalar-meson family that is related to the Higgs boson of the hadronic sector.
This is a result of principle, because the schemes of the nonlinear
realization of the chiral symmetry have been considered which do without the
Higgs mesons. One can think that a linear realization of the chiral symmetry
(at least, for the lightest states and related phenomena) is valid. First,
this is a simple and beautiful mechanism that works also in other fields of
physics, for example, in superconductivity. Second, the effective Lagrangians
obtained on the basis of this mechanism (the Nambu -- Jona-Lasinio and other
models) describe perfectly the ground states and related phenomena. The only
weak link of this approach was the absence of the $\sigma$-meson below 1 GeV.

Note also that the $f_0 (665)$ changes but does not solve the problem of
unusual properties of the scalar mesons that prevent the scalar nonet to be
made up.

Let us also notice that the character of the $f_0(665)$ pole-cluster (namely,
a considerable shift of the pole on sheet III towards the imaginary axis) can
point to the unconsidered channel with which this state is, possibly, coupled
strongly, and the threshold of which is situated below 600 MeV. In this energy
region, only one channel is opened: this is the $4\pi$ channel. It is
interesting to verify this assumption, because it concerns such an important
state.

This analysis does not reveal the ${f_0}(1370)$ resonance; therefore, if
this meson exists, it must be weakly coupled with the $\pi\pi$ channel,
{\it i.e.} be described by the pole cluster of the type ({\bf b}) (this would
testify to the dominant $s{\bar s}$ component in this state; as to that
assignment of the ${f_0}(1370)$ resonance, we agree, {\it e.g.,} with the
work \cite{Shakin1}).

The $f_0 (1500)$ state is represented by the pole cluster on the Riemann
surface of the $S$-matrix of the type ({\bf c}) which corresponds to a
glueball. This type of cluster ({\it i.e.} the presence of the zeros,
corresponding to the state, on the physical sheet of both $\pi\pi$ and
$K\overline{K}$ scattering) reflects the flavour-singlet structure of the
glueball wave-function and is only a necessary condition of the glueball
nature of the $f_0 (1500)$ state.
Let us also pay attention to the strong coupling of the $f_0(1500)$ state
with the $\pi\pi$ system, and to that in the model-independent approach, one
can obtain a qualitative indication -- how much is the admixture of other
states ($q\bar q,q{\bar q}g$, etc.)? To this end, one must consider the
$\pi\pi\to K\overline{K}$ process in our 3-channel approach \cite{KMS-nc96}
and determine the coupling constants of the $f_0(1500)$ with the other members
of the pseudoscalar nonet.

We emphasize that the obtained results are model-independent, since they are
based on the first principles and on the mathematical fact that a local
behaviour of analytic functions, determined on the Riemann surface, is
governed by the nearest singularities on all sheets.

We think that multichannel states are most adequately represented by clusters,
{\it i.e.} by the pole positions on all corresponding sheets. The pole
positions are rather stable characteristics for various models, whereas masses
and widths are very model-dependent for wide resonances.

Finally, note that in the model-independent approach, there are many adjusted
parameters (although, {\it e.g.} for the $\pi\pi$ scattering, they all are
positions of poles describing resonances). The number of these parameters can
be diminished by some dynamic assumptions, but this is another approach and of
other value.

\section*{Acknowledgments}
The authors are grateful to S.~Dubni{\'{c}}ka, S.B.~Gerasimov,
V.A.~Meshcheryakov, V.N.~Pervushin, A.I.~Titov, M.K.~Volkov and V.L.~Yudichev 
for useful discussions and interest in this work.

This work has been supported by the Grant Program of Plenipotentiary of Slovak
Republic at JINR.  Yu.S. and M.N. were supported in part by the Slovak
Scientific Grant Agency, Grant VEGA No. 2/7175/20; and D.K., by Grant VEGA
No. 2/5085/99. Yu.S. is grateful to the Bogoliubov-Infeld Program for giving
the travel grant to participate in the MESON'2000 Workshop.

\underline{Figure Captions}\\

\noindent
Fig.1: Uniformization plane for the \protect$\pi\pi$-scattering amplitude.
The Roman numerals (I,\ldots,IV) denote the images of the corresponding sheets
of the Riemann surface; the thick line represents the physical region (the
points i, 1 and b correspond to the \protect$\pi\pi, K\overline{K}$ thresholds
and \protect$s=\infty)$, respectively); the shaded lines are the images of the
corresponding edges of the left-hand cut. The depicted positions of poles
\protect($*$) and of zeros \protect($\circ$) give the representation of the
type ({\bf a}) resonance in \protect$S_{11}$. \\
\noindent
Fig.2: The energy dependence of the phase shift \protect($\delta_{1}$) of the
\protect$\pi\pi$-scattering amplitude obtained on the basis of a simultaneous
analysis of the experimental data on the coupled processes
\protect$\pi\pi\to \pi\pi,K\overline{K}$ in the channel with
\protect$I^GJ^{PC}=0^+0^{++}$. The data on the \protect$\pi\pi$ scattering
are taken from Refs.\cite{Hyams}.\\
\noindent
Fig.3: The same as in Fig.2 but for the elasticity parameter \protect$\eta$.\\
\noindent
Fig.4: The energy dependence of the \protect($|S_{12}|$) obtained on the
basis of a simultaneous analysis of the experimental data on the coupled
processes \protect$\pi\pi\to \pi\pi,K\overline{K}$ in the channel with
\protect$I^GJ^{PC}=0^+0^{++}$. The data on the process
\protect$\pi\pi\to K\overline{K}$ are taken from Ref.\cite{Wickl}.\\
\noindent
Fig.5: The same as in Fig.4 but for the phase shift \protect($\delta_{12}$).

\newpage

\begin{figure}[ht]
\vskip 1.cm
\centerline{\epsfxsize=.6 \hsize \epsffile{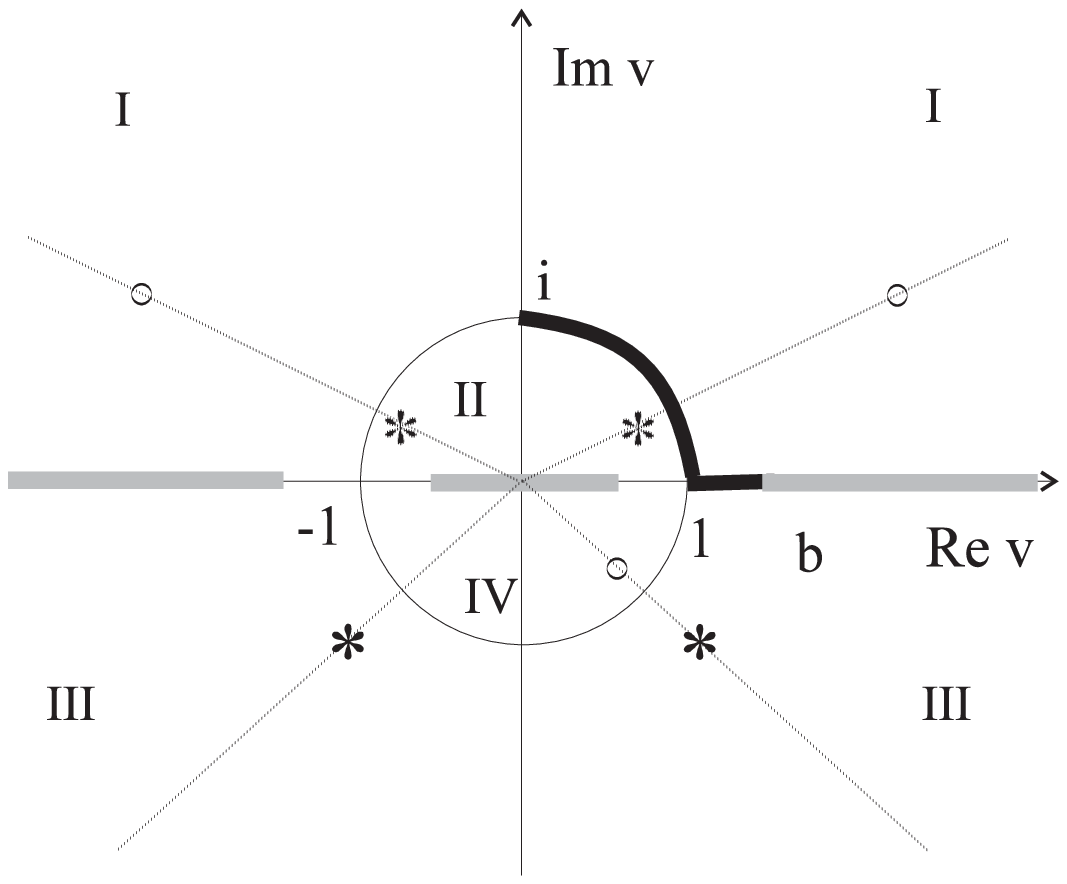}}
\vskip -.1cm
\caption{}
\end{figure}

\begin{figure}[ht]
\vskip 0.5cm
\centerline{\epsfxsize=.7 \hsize \epsffile{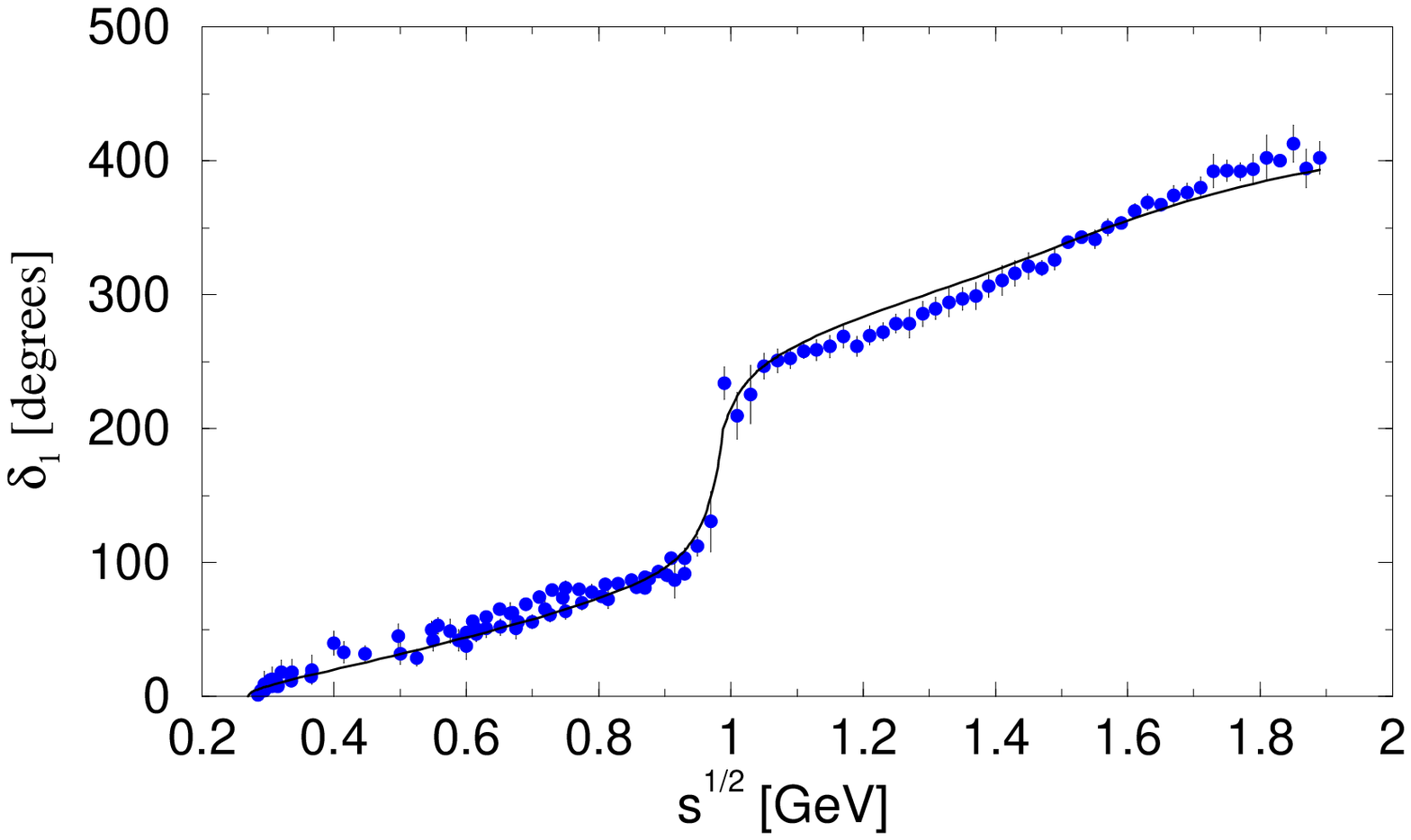}}
\vskip -.1cm
\caption{}
\end{figure}

\begin{figure}[ht]
\vskip 0.5cm
\centerline{\epsfxsize=.7 \hsize \epsffile{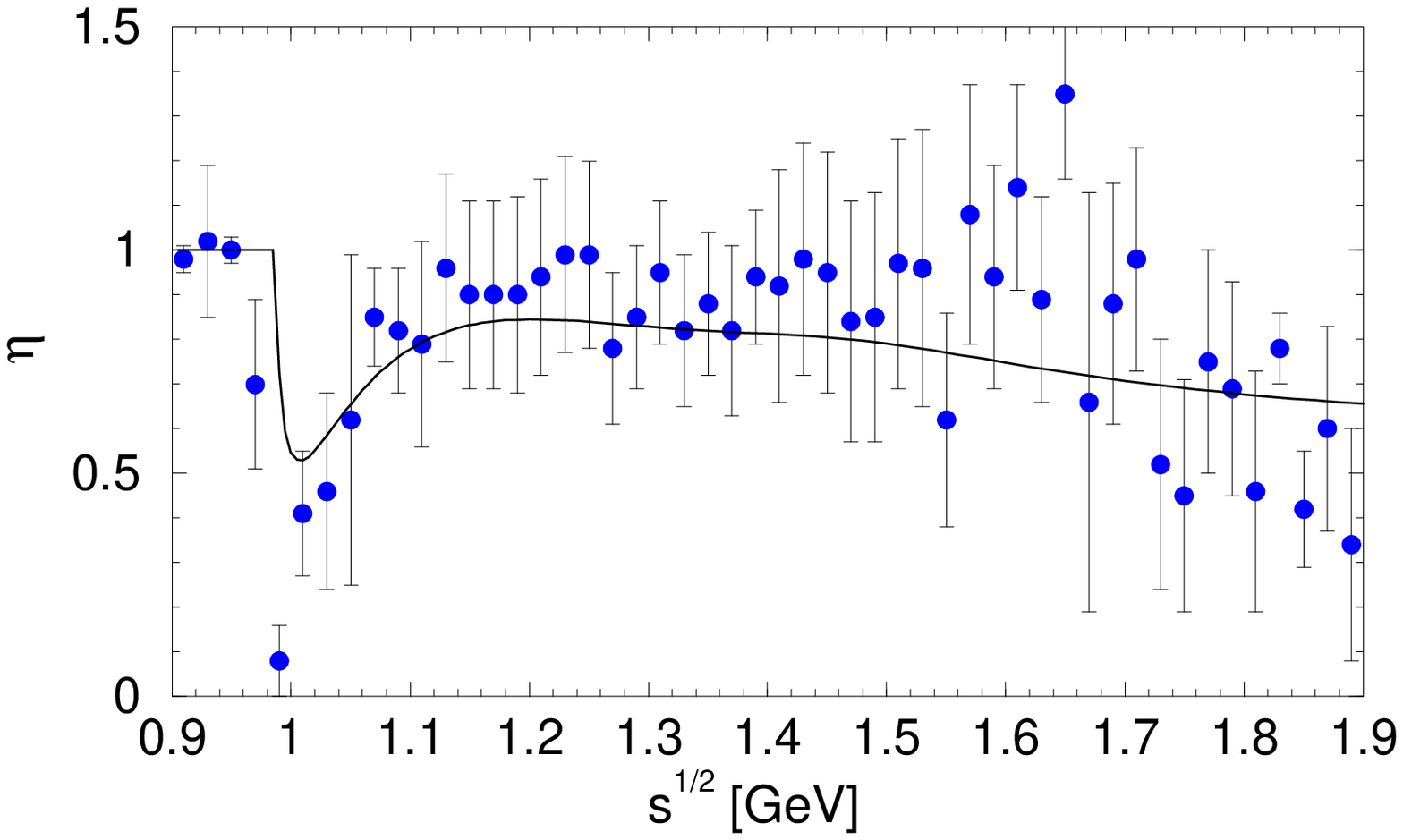}}
\vskip -.1cm
\caption{}
\end{figure}

\begin{figure}[ht]
\vskip 0.1cm
\centerline{\epsfxsize=.7 \hsize \epsffile{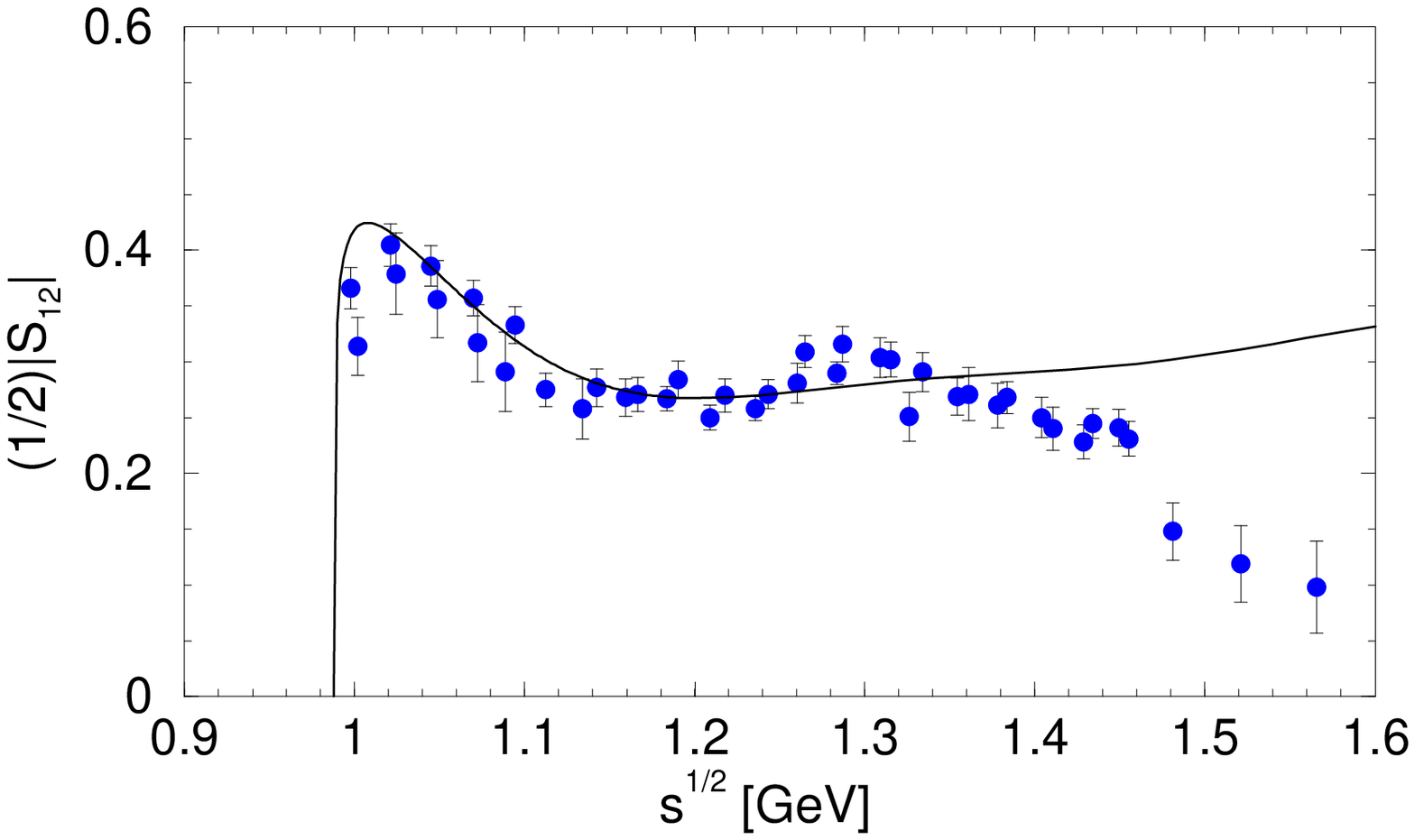}}
\vskip -.1cm
\caption{}
\end{figure}

\begin{figure}[ht]
\vskip 0.5cm
\centerline{\epsfxsize=.7 \hsize \epsffile{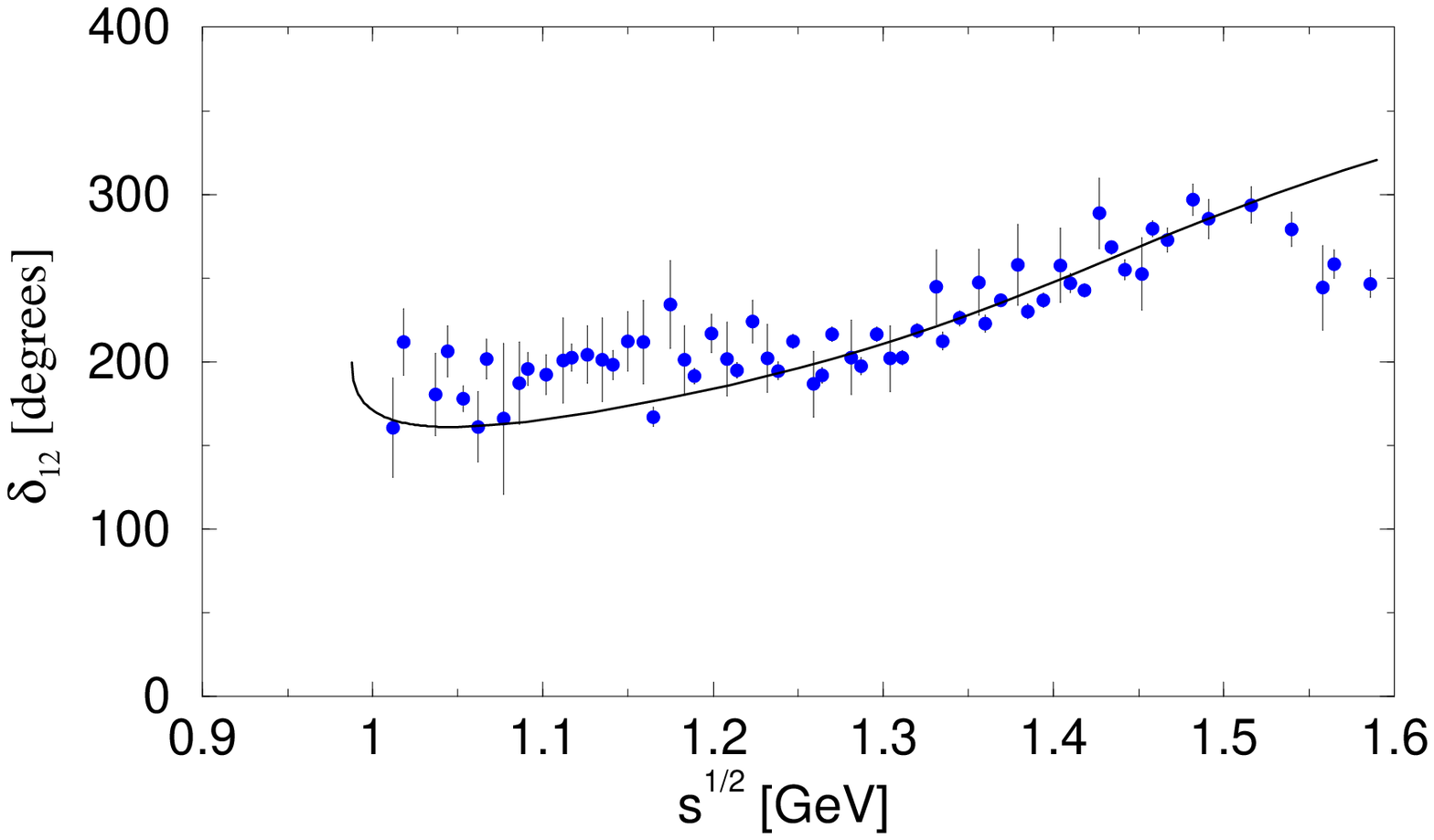}}
\vskip -.1cm
\caption{}
\end{figure}


\begin{thebibliography}{30}
\bibitem{KMS-nc96} D. Krupa, V.A. Meshcheryakov, Yu.S. Surovtsev,
Nuovo Cimento {\bf 109 A}, 281 (1996).
\bibitem{PDG-98} Review of Particle Physics, Europ. Phys. J. {\bf C3}, 1
(1998).
\bibitem{NJL} Y. Nambu, G. Jona-Lasinio, Phys. Rev. {\bf 122}, 345 (1961);
M.K. Volkov, Ann. Phys. {\bf 157}, 282 (1984); T. Hatsuda, T. Kunihiro,
Phys. Rep. {\bf 247}, 223 (1994); R. Delbourgo, M.D. Scadron, Mod. Phys. Lett.
{\bf A10}, 251 (1995).
\bibitem{Zou} B.S. Zou, D.V. Bugg,  Phys. Rev. {\bf D48}, R3948 (1993);
{\bf D50}, 591 (1994).
\bibitem{Svec} M. Svec, Phys. Rev. {\bf D53}, 2343 (1996).
\bibitem{Ishida} S. Ishida et al., Progr. Theor. Phys. {\bf 95}, 745 (1996);
{\bf 98}, 621 (1997).
\bibitem{Torn1} N. T\"ornqvist, Phys. Rev. Lett. {\bf 76}, 1575 (1996).
\bibitem{Kaminski} R. Kami\'nski, L. Le\'sniak, B. Loiseau, Eur. Phys. J.
C 9 (1999) 141.
\bibitem{KMS-cz88} D. Krupa, V.A. Meshcheryakov, Yu.S. Surovtsev, Yad. Fiz.
{\bf 43}, 231 (1986); Czech. J. Phys. {\bf B38}, 1129 (1988).
\bibitem{KMS-L86} D. Krupa, V.A. Meshcheryakov, Yu.S. Surovtsev,
{\it Pole-Cluster Representations as Tests of Multichannel Resonances and
Coupled Processes in Scalar Sector}. Proc. Intern. Conf."HADRON STRUCTURE'96"
(Star\'a Lesn\'a, Slovak Republic, 12-16 February 1996), ed. L.Martinovi\v{c}
and P.Stri\v{z}enec, Dubna, 1996, p.86.
\bibitem{MP-92} D. Morgan, M.R. Pennington, Phys. Rev. {\bf D48},1185 (1993).
\bibitem{LN} K.J. Le Couteur, Proc. Roy. Soc. {\bf A256}, 115 (1960);
R.G. Newton, J. Math. Phys. {\bf 2}, 188 (1961).
\bibitem{Kato} M. Kato, Ann. Phys. {\bf 31}, 130 (1965).
\bibitem{Bohacik} J. Bohacik, H. K\"uhnelt, Phys. Rev. {\bf D21}, 1342
(1980).
\bibitem{Hyams} B. Hyams et al., Nucl. Phys. {\bf B64}, 134 (1973); ibid.,
{\bf B100}, 205 (1975).
A. Zylbersztejn et al., Phys. Lett. {\bf B38}, 457 (1972).
P. Sonderegger, P. Bonamy, Proc. 5th Intern. Conf. on Elementary Particles,
Lund, 1969, paper 372. J.R. Bensinger et al., Phys. Lett. {\bf B36}, 134
(1971). J.P. Baton et al., Phys. Lett. {\bf B33}, 525, 528 (1970).
P. Baillon et al., Phys. Lett. {\bf B38}, 555 (1972). L. Rosselet et al.,
Phys. Rev. {\bf D15}, 574 (1977). A.A. Kartamyshev et al., Pis'ma v Zh. Eksp.
Teor. Fiz. {\bf 25}, 68 (1977). A.A. Bel'kov et al., Pis'ma v Zh. Eksp. Teor.
Fiz. {\bf 29}, 652 (1979).
\bibitem{Wickl} A.B. Wicklund et al., Phys. Rev. Lett. {\bf 45}, 1469 (1980).
D. Cohen et al., Phys. Rev. {\bf D22}, 2595 (1980).
A. Etkin et al., Phys. Rev. {\bf D25}, 1786 (1982).
\bibitem{skn-99} Yu.S. Surovtsev, D. Krupa, M.Nagy, {\it e-Print Archive}:
hep-ph/0005090 (2000).
\bibitem{Weinberg} S. Weinberg, Phys. Rev. Lett. {\bf 17}, 616 (1966).
B.W. Lee, H.T. Nieh, Phys. Rev. {\bf 166}, 1507 (1968).
\bibitem{Gasser} J. Gasser, H. Leutwyler, Ann. Phys. {\bf 158}, 142 (1984).
\bibitem{Bijnens} J. Bijnens et al., Phys. Lett. {\bf B374}, 210 (1996).
\bibitem{Volkov} M.K. Volkov, Physics of Elementary Particles and Atomic
Nuclei, v.17, part 3, 433 (1986).
\bibitem{Volk-Yud} M.K. Volkov, V.L. Yudichev, M. Nagy, Nuovo Cimento
{\bf 112 A}, 225 (1999).
\bibitem{Shakin1} L.S. Celenza, Huangsheng Wang, C.M. Shakin, {\it Preprint of
Brooklyn College of the City Univ. of New York}, BCCNT: 00/041/289 (2000).

\end{thebibliography}
\end{document}